\documentclass[11pt]{article}
\usepackage[a4paper,margin=1in]{geometry}
\usepackage{amsmath,amssymb,amsthm}
\usepackage[hypertexnames=false]{hyperref}
\usepackage{graphicx}
\usepackage{booktabs}
\usepackage{url}
\usepackage{enumitem}
\usepackage{microtype}
\usepackage{algorithm}
\usepackage{algorithmic}
\usepackage{tikz}
\usetikzlibrary{shapes.geometric, arrows.meta, positioning, fit}

\newtheorem{theorem}{Theorem}[section]
\newtheorem{definition}[theorem]{Definition}

\title{Constant-Size Cryptographic Evidence Structures for Regulated AI Workflows}

\author{
  Leo Kao \\
  Codebat Technologies Inc. \\
  \texttt{leo@codebat.ai}
}

\date{\today}

\begin{document}
\maketitle

\begin{abstract}
Regulated AI workflows---such as those used in clinical trials, medical decision support, and financial compliance---must satisfy strict auditability and integrity requirements. Existing logging and audit-trail mechanisms typically rely on variable-length records, bulky cryptographic transcripts, or ad-hoc database schemas. These designs suffer from metadata leakage, irregular performance characteristics, and weak alignment with formal security notions and regulatory semantics.

This paper introduces \emph{constant-size cryptographic evidence structures}, a general abstraction for representing verifiable audit evidence for AI workflows in regulated environments. Each evidence item is a fixed-size tuple of cryptographic fields, designed to (i) provide strong binding to workflow events and configurations, (ii) support constant-size storage and uniform verification cost per event, and (iii) compose cleanly with hash-chain and Merkle-based audit constructions. We formalize a simple model of regulated AI workflows, define syntax and algorithms for evidence structures, and formalize security properties---evidence binding, tamper detection, and non-equivocation---via game-based definitions, proving that our construction satisfies them under standard assumptions (collision-resistant hashing and EUF-CMA signatures).

We present a generic hash-and-sign construction that instantiates this abstraction using a collision-resistant hash function and a standard digital signature scheme. We then show how to integrate the construction with hash-chained logs, Merkle-tree anchoring, and (optionally) trusted execution environments, and we analyze the asymptotic complexity of evidence generation and verification. Finally, we implement a prototype library and report microbenchmark results on commodity hardware, demonstrating that the per-event overhead of constant-size evidence is small and predictable. The design is informed by industrial experience with regulated AI systems at Codebat Technologies Inc., while the paper focuses on the abstraction, algorithms, and their security and performance characteristics. This work aims to provide a foundation for standardized audit mechanisms in regulated AI, with implications for clinical trial management, pharmaceutical compliance, and medical AI governance.
\end{abstract}


\section{Introduction}

AI systems deployed in regulated domains---including clinical trials, medical imaging, drug safety monitoring, and financial compliance---must provide strong guarantees of \emph{traceability}, \emph{integrity}, and \emph{accountability}. Regulators and institutional review boards (IRBs) require that each critical operation (e.g., data access, model inference, randomization, or protocol change) leave a verifiable audit trail that can be inspected months or years later. At the same time, operators must protect sensitive information such as personal health information (PHI), internal model details, or proprietary algorithms.

In practice, most audit trails are built from one or more of the following components:
\begin{itemize}[leftmargin=*]
  \item Variable-length application logs (e.g., \texttt{syslog}-style entries),
  \item Ad-hoc database schemas updated by application code,
  \item Blockchain or distributed-ledger events with free-form payloads.
\end{itemize}
These mechanisms are flexible but misaligned with the needs of regulated AI workflows. Variable-length records leak metadata (e.g., larger records may correspond to imaging data or long clinical notes), incur non-uniform storage and verification costs, and complicate high-throughput verification. Ledger-based solutions often incur unpredictable gas or transaction fees and push complexity into the smart-contract layer.

Secure logging and audit-trail systems have been studied extensively~\cite{schneier1999secure, crosby2009efficient, bellare1997forward}, as have hash-chain and Merkle-tree-based data structures~\cite{merkle1980protocols, haber1991time}. More recently, blockchain-based transparency logs~\cite{laurie2013certificate, nakamoto2008bitcoin} and trusted execution environments (TEEs)~\cite{costan2016intel, sabt2015ted} have been used to build tamper-evident records for security-critical operations. However, these approaches typically operate on variable-length or application-specific records and do not treat a fixed-size, hardware-friendly evidence representation as a first-class abstraction.

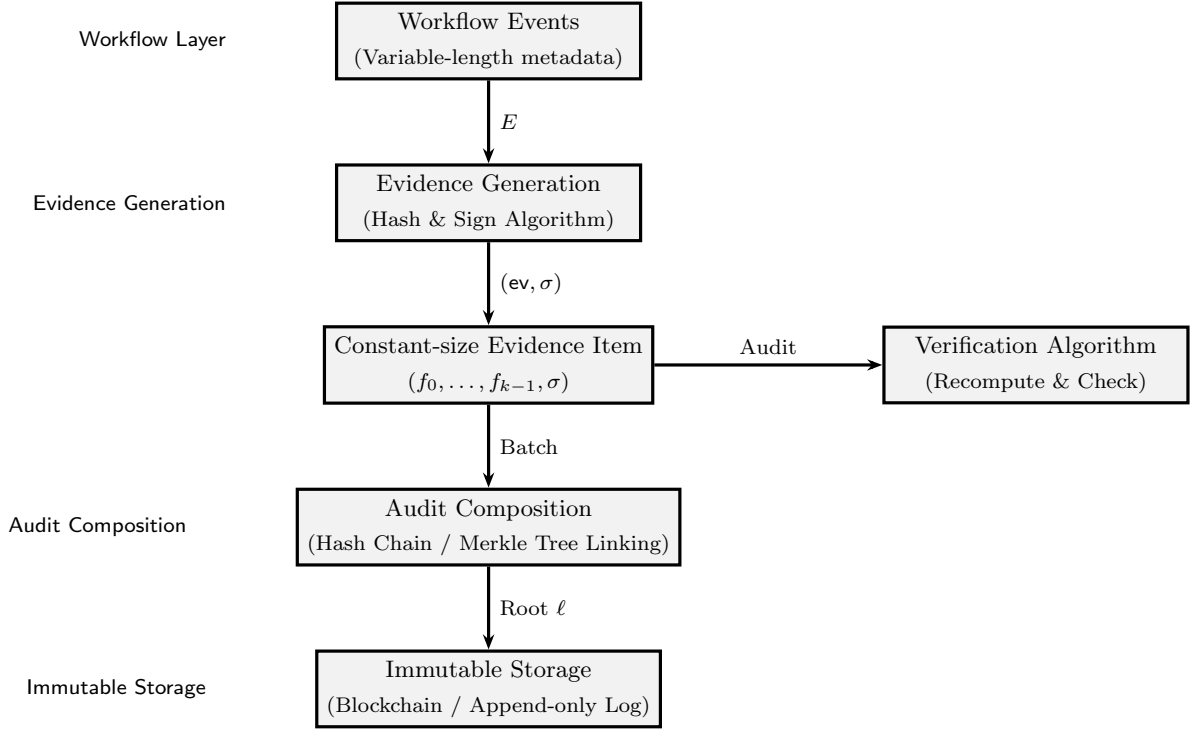
\begin{figure}[t]
\centering
\begin{tikzpicture}[
  node distance=1.1cm,
  box/.style={
    rectangle,
    draw,
    very thick,
    minimum width=4.0cm,
    minimum height=0.9cm,
    align=center,
    font=\footnotesize
  },
  arrow/.style={
    -{Stealth[length=2mm]},
    very thick
  },
  layer/.style={
    font=\scriptsize\sffamily
  }
]

\node[box, fill=gray!10] (events) {Workflow Events\\[2pt]\scriptsize(Variable-length metadata)};

\node[box, fill=gray!10, below=of events] (generate) {Evidence Generation\\[2pt]\scriptsize(Hash \& Sign Algorithm)};

\node[box, fill=gray!10, below=of generate] (evidence) {Constant-size Evidence Item\\[2pt]\scriptsize$(f_0, \ldots, f_{k-1}, \sigma)$};

\node[box, fill=gray!10, below=of evidence] (link) {Audit Composition\\[2pt]\scriptsize(Hash Chain / Merkle Tree Linking)};

\node[box, fill=gray!10, below=of link] (anchor) {Immutable Storage\\[2pt]\scriptsize(Blockchain / Append-only Log)};

\node[box, fill=gray!10, right=3cm of evidence] (verify) {Verification Algorithm\\[2pt]\scriptsize(Recompute \& Check)};

\draw[arrow] (events) -- (generate) node[midway, right, font=\scriptsize] {$E$};
\draw[arrow] (generate) -- (evidence) node[midway, right, font=\scriptsize] {$(\mathsf{ev}, \sigma)$};
\draw[arrow] (evidence) -- (link) node[midway, right, font=\scriptsize] {Batch};
\draw[arrow] (link) -- (anchor) node[midway, right, font=\scriptsize] {Root $\ell$};

\draw[arrow] (evidence.east) -- (verify.west) node[midway, above, font=\scriptsize] {Audit};

\node[layer, left=1.3cm of events]  {Workflow Layer};
\node[layer, left=1.3cm of generate] {Evidence Generation};
\node[layer, left=1.3cm of link] {Audit Composition};
\node[layer, left=1.3cm of anchor] {Immutable Storage};

\end{tikzpicture}
\caption{Overview of the evidence workflow from workflow events to immutable storage and verification.}
\label{fig:evidence-workflow}
\end{figure}

\paragraph{Our approach.}
We propose to view audit evidence for regulated AI workflows as \emph{constant-size cryptographic structures} generated and verified by well-defined algorithms. Each evidence item is a tuple of a fixed number of fields, where each field is a fixed-length bitstring (e.g., the output of a hash function). Conceptually, each item binds to:
\begin{itemize}[leftmargin=*]
  \item A well-defined unit of work (e.g., a model invocation or randomization step),
  \item The relevant inputs and outputs (via commitments),
  \item The execution environment (e.g., an environment identifier or attestation digest),
  \item A link into a larger audit structure (e.g., a hash chain or Merkle root),
  \item A cryptographic authenticator that binds all fields together.
\end{itemize}
The central requirement is that the \emph{representation size} of each evidence item is independent of the size or type of the underlying event, which enables constant-size storage and uniform verification cost per event. Figure~\ref{fig:evidence-workflow} illustrates the high-level architecture of the evidence structure system.

\paragraph{Contributions.}
We make the following contributions:
\begin{enumerate}[leftmargin=*]
  \item We formalize a simple model of regulated AI workflows and define the syntax of \emph{constant-size evidence structures}, including generation and verification algorithms.
  \item We identify design objectives for such structures, including fixed-size representation, strong cryptographic binding, efficient batch verification, and compatibility with hash-chain and Merkle-based audit frameworks.
  \item We present a generic hash-and-sign construction that instantiates the abstraction using a collision-resistant hash function and a digital signature scheme.
  \item We formalize three security properties (evidence binding, tamper detection, non-equivocation) via game-based definitions and prove that each holds under standard assumptions with tight reductions.
  \item We discuss how the construction composes with hash-chained logs, Merkle-tree anchoring, and optional TEE-based environment binding, and analyze the asymptotic complexity of evidence generation and verification.
  \item We implement a prototype library and report microbenchmark results that illustrate the performance characteristics of the abstraction on commodity hardware.
\end{enumerate}

\section{Model and Design Objectives}

\subsection{Workflow Model}

We consider a workflow as a sequence (or directed acyclic graph) of \emph{events}. Each event $E$ has:
\begin{itemize}[leftmargin=*]
  \item a unique identifier $\mathsf{id}(E)$,
  \item structured metadata $\mathsf{meta}(E)$ (e.g., actor, time, configuration),
  \item a set of input references $\mathsf{in}(E)$,
  \item a set of output references $\mathsf{out}(E)$.
\end{itemize}
We write $M(E)$ for the full metadata associated with $E$, including these components. We do not assume any particular domain (e.g., clinical, financial); the abstraction is intended to be domain-agnostic.

\subsection{Adversary and Environment}

We assume an adversary who may:
\begin{itemize}[leftmargin=*]
  \item Control or compromise some application components that generate events,
  \item Observe or tamper with storage and network channels used to store and transmit audit data,
  \item Attempt to suppress, insert, or modify audit evidence,
  \item Attempt to correlate audit metadata with sensitive events.
\end{itemize}
We assume the existence of standard cryptographic primitives (hash functions, signature schemes) with their usual security properties, and we treat external anchoring mechanisms (e.g., append-only logs or blockchains) as black-box components that can store short digests.

\subsection{Design Objectives}

We highlight several objectives that guide our abstraction.

\paragraph{Fixed-size representation.}
Each evidence item associated with an event $E$ must have the same size, independent of the structure or length of $M(E)$. The number of fields and their bit-lengths are fixed at system setup time.

\paragraph{Cryptographic binding.}
An evidence item should cryptographically bind to the salient aspects of $E$, including its inputs, outputs, configuration, and (optionally) execution environment. Any modification to these aspects should require either finding hash collisions or forging signatures to avoid detection.

\paragraph{Efficient batch verification.}
Evidence items should be verifiable independently and in parallel, with verification cost that is uniform across items and dominated by a small number of hash computations and signature verifications.

\paragraph{Compatibility with audit structures.}
Evidence items should integrate cleanly with existing audit structures such as hash chains and Merkle trees~\cite{merkle1980protocols}, so that large collections of items can be anchored succinctly.

\paragraph{Regulator-aligned semantics.}
The fields in an evidence item should support straightforward mapping to regulatory concepts such as actors, timestamps, model versions, and policy states, without hard-coding any particular regulation.

\section{Evidence Structures: Syntax and Algorithms}

\subsection{Syntax}

Let $\lambda$ be a security parameter. A \emph{constant-size evidence structure} for workflows is defined by:
\begin{itemize}[leftmargin=*]
  \item A finite index set $I = \{0,1,\dots,k-1\}$ for some fixed $k \in \mathbb{N}$,
  \item A field length parameter $\lambda$,
  \item A type of evidence items:
  \[
    \mathsf{Ev} = \{0,1\}^{\lambda} \times \dots \times \{0,1\}^{\lambda}
  \]
  consisting of $k$ fields $(f_0,\dots,f_{k-1})$, each a $\lambda$-bit string.
\end{itemize}

The interpretation of $f_i$ is application-specific but must be fixed by the system designer. Typical roles include:
\begin{itemize}[leftmargin=*]
  \item context commitments (e.g., to workflow identifiers),
  \item commitments to inputs or outputs,
  \item environment identifiers or attestations,
  \item links to other evidence items or audit structures,
  \item signature- or key-related material.
\end{itemize}

\subsection{Algorithms}

An evidence structure is equipped with the following algorithms:

\begin{itemize}[leftmargin=*]
  \item $\mathsf{Setup}(1^\lambda) \rightarrow \mathsf{pp}$: generates public parameters $\mathsf{pp}$ (e.g., hash function description, field interpretation).
  \item $\mathsf{KeyGen}(1^\lambda) \rightarrow (sk, pk)$: generates signing and verification keys for the entity that produces evidence.
  \item $\mathsf{Generate}(\mathsf{pp}, sk, E) \rightarrow (\mathsf{ev}, \sigma)$: given an event $E$ and secret key $sk$, produces an evidence item $\mathsf{ev} \in \mathsf{Ev}$ and an authenticator $\sigma$ (e.g., a signature).
  \item $\mathsf{Verify}(\mathsf{pp}, pk, E, \mathsf{ev}, \sigma) \rightarrow \{\mathsf{accept}, \mathsf{reject}\}$: checks whether $\mathsf{ev}$ and $\sigma$ are valid for event $E$ under public key $pk$.
  \item $\mathsf{Link}(\mathsf{pp}, S) \rightarrow \ell$: given a set or sequence $S$ of evidence items, produces a link value $\ell$ (e.g., a hash-chain tip or Merkle root).
\end{itemize}

\section{A Generic Hash-and-Sign Construction}
\label{sec:construction}

\subsection{Components}

Let $H: \{0,1\}^* \rightarrow \{0,1\}^{\lambda}$ be a collision-resistant hash function, and let $(\mathsf{KeyGen}^{\mathsf{sig}}, \mathsf{Sign}, \mathsf{Verify}^{\mathsf{sig}})$ be a digital signature scheme that is existentially unforgeable under chosen-message attacks (EUF-CMA)~\cite{goldwasser1988digital, nistfips186}. We assume that the system designer provides a family of deterministic encoding functions
\[
  \phi_i : \mathcal{E} \rightarrow \{0,1\}^*
\]
for each index $i \in I$, where $\mathcal{E}$ is the set of events. We require that the encoding family $\{\phi_i\}_{i \in I}$ is \emph{collectively injective}: for any two distinct events $E \neq E'$, there exists at least one index $i \in I$ such that $\phi_i(E) \neq \phi_i(E')$. This ensures that distinct events produce differing inputs to the hash function in at least one field, which is essential for the binding property established in Section~\ref{sec:security}.

\subsection{Algorithms in Pseudocode}

\begin{algorithm}[t]
\caption{$\mathsf{Generate}(\mathsf{pp}, sk, E)$}
\label{alg:generate}
\begin{algorithmic}[1]
  \STATE Parse $\mathsf{pp}$ to obtain $H$, index set $I$, encodings $\{\phi_i\}_{i \in I}$
  \FOR{each $i \in I$}
    \STATE $m_i \leftarrow \phi_i(E)$ \COMMENT{encode event $E$ for field $i$}
    \STATE $f_i \leftarrow H(m_i)$
  \ENDFOR
  \STATE $\mathsf{ev} \leftarrow (f_0, f_1, \dots, f_{k-1})$
  \STATE $\sigma \leftarrow \mathsf{Sign}_{sk}(\mathsf{ev})$
  \STATE \textbf{return} $(\mathsf{ev}, \sigma)$
\end{algorithmic}
\end{algorithm}

\begin{algorithm}[t]
\caption{$\mathsf{Verify}(\mathsf{pp}, pk, E, \mathsf{ev}, \sigma)$}
\label{alg:verify}
\begin{algorithmic}[1]
  \STATE Parse $\mathsf{pp}$ to obtain $H$, index set $I$, encodings $\{\phi_i\}_{i \in I}$
  \STATE Parse $\mathsf{ev}$ as $(f_0, f_1, \dots, f_{k-1})$
  \FOR{each $i \in I$}
    \STATE $m_i \leftarrow \phi_i(E)$
    \STATE $f'_i \leftarrow H(m_i)$
    \IF{$f'_i \neq f_i$}
      \STATE \textbf{return} $\mathsf{reject}$
    \ENDIF
  \ENDFOR
  \IF{$\mathsf{Verify}^{\mathsf{sig}}_{pk}(\mathsf{ev}, \sigma) = 0$}
    \STATE \textbf{return} $\mathsf{reject}$
  \ENDIF
  \STATE \textbf{return} $\mathsf{accept}$
\end{algorithmic}
\end{algorithm}

\begin{algorithm}[t]
\caption{$\mathsf{Link}( \mathsf{pp}, S )$ via hash chaining}
\label{alg:link}
\begin{algorithmic}[1]
  \STATE $\ell_0 \leftarrow 0^\lambda$ \COMMENT{all-zero initialization}
  \STATE Let $S = (\mathsf{ev}_1, \dots, \mathsf{ev}_n)$ be an ordered sequence
  \FOR{$j = 1$ \TO $n$}
    \STATE $\ell_j \leftarrow H(\ell_{j-1} \parallel \mathsf{ev}_j)$
  \ENDFOR
  \STATE \textbf{return} $\ell_n$
\end{algorithmic}
\end{algorithm}

\section{Security Analysis}
\label{sec:security}

We now formalize the security guarantees of the hash-and-sign evidence construction. We define three security properties via game-based experiments~\cite{bellarerogaway1993, bonehshoup} and prove that the construction satisfies each under standard cryptographic assumptions~\cite{katzlindell}: collision resistance of the hash function~\cite{rogaway2004cryptographic} and existential unforgeability of the signature scheme~\cite{goldwasser1988digital}. Throughout this section, $\lambda$ denotes the security parameter, $H\colon\{0,1\}^*\to\{0,1\}^\lambda$ denotes a collision-resistant hash function, and $(\mathsf{KeyGen}^{\mathsf{sig}},\mathsf{Sign},\mathsf{Verify}^{\mathsf{sig}})$ denotes an EUF-CMA-secure signature scheme.

\subsection{Security Definitions}

We define three game-based experiments that capture the core security goals of constant-size evidence structures.

\begin{definition}[Evidence Binding]
\label{def:evidence-binding}
Let $\Pi$ be an evidence structure with field count $k$. The \emph{evidence-binding} experiment $\mathbf{Exp}^{\mathrm{Bind}}_{\Pi,\mathcal{A}}(\lambda)$ proceeds as follows:
\begin{enumerate}
  \item Run $\mathsf{pp} \leftarrow \mathsf{Setup}(1^\lambda)$ and $(sk, pk) \leftarrow \mathsf{KeyGen}(1^\lambda)$.
  \item The adversary $\mathcal{A}$ receives $\mathsf{pp}$ and $pk$, and has oracle access to $\mathsf{Generate}(\mathsf{pp}, sk, \cdot)$.
  \item $\mathcal{A}$ outputs $(E, E', \mathsf{ev}, \sigma)$.
  \item The experiment outputs $1$ if and only if $E \neq E'$, $\mathsf{Verify}(\mathsf{pp}, pk, E, \mathsf{ev}, \sigma) = \mathsf{accept}$, and $\mathsf{Verify}(\mathsf{pp}, pk, E', \mathsf{ev}, \sigma) = \mathsf{accept}$.
\end{enumerate}
The \emph{evidence-binding advantage} of $\mathcal{A}$ is $\mathbf{Adv}^{\mathrm{Bind}}_{\Pi,\mathcal{A}}(\lambda) = \Pr[\mathbf{Exp}^{\mathrm{Bind}}_{\Pi,\mathcal{A}}(\lambda) = 1]$.
\end{definition}

\begin{definition}[Tamper Detection --- Unforgeability]
\label{def:tamper-detection}
Let $\Pi$ be an evidence structure. The \emph{tamper-detection} experiment $\mathbf{Exp}^{\mathrm{TD}}_{\Pi,\mathcal{A}}(\lambda)$ proceeds as follows:
\begin{enumerate}
  \item Run $\mathsf{pp} \leftarrow \mathsf{Setup}(1^\lambda)$ and $(sk, pk) \leftarrow \mathsf{KeyGen}(1^\lambda)$.
  \item The adversary $\mathcal{A}$ receives $\mathsf{pp}$ and $pk$, and has oracle access to $\mathsf{Generate}(\mathsf{pp}, sk, \cdot)$. Let $\mathcal{Q}$ be the set of events queried to $\mathsf{Generate}$.
  \item $\mathcal{A}$ outputs $(E^*, \mathsf{ev}^*, \sigma^*)$.
  \item The experiment outputs $1$ if and only if $E^* \notin \mathcal{Q}$ and $\mathsf{Verify}(\mathsf{pp}, pk, E^*, \mathsf{ev}^*, \sigma^*) = \mathsf{accept}$.
\end{enumerate}
The \emph{tamper-detection advantage} of $\mathcal{A}$ is $\mathbf{Adv}^{\mathrm{TD}}_{\Pi,\mathcal{A}}(\lambda) = \Pr[\mathbf{Exp}^{\mathrm{TD}}_{\Pi,\mathcal{A}}(\lambda) = 1]$.
\end{definition}

\begin{definition}[Non-Equivocation --- Chain Integrity]
\label{def:non-equivocation}
Let $\Pi$ be an evidence structure with hash-chain linking (Algorithm~\ref{alg:link}). The \emph{non-equivocation} experiment $\mathbf{Exp}^{\mathrm{NE}}_{\Pi,\mathcal{A}}(\lambda)$ proceeds as follows:
\begin{enumerate}
  \item Run $\mathsf{pp} \leftarrow \mathsf{Setup}(1^\lambda)$.
  \item The adversary $\mathcal{A}$ receives $\mathsf{pp}$.
  \item $\mathcal{A}$ outputs two sequences of evidence items $S = (\mathsf{ev}_1, \dots, \mathsf{ev}_n)$ and $S' = (\mathsf{ev}'_1, \dots, \mathsf{ev}'_m)$.
  \item The experiment outputs $1$ if and only if $S \neq S'$ and $\mathsf{Link}(\mathsf{pp}, S) = \mathsf{Link}(\mathsf{pp}, S')$.
\end{enumerate}
The \emph{non-equivocation advantage} of $\mathcal{A}$ is $\mathbf{Adv}^{\mathrm{NE}}_{\Pi,\mathcal{A}}(\lambda) = \Pr[\mathbf{Exp}^{\mathrm{NE}}_{\Pi,\mathcal{A}}(\lambda) = 1]$.
\end{definition}

\paragraph{Remark.}
Evidence binding (Definition~\ref{def:evidence-binding}) is analogous to the binding property of cryptographic commitment schemes: a single evidence item cannot serve as valid evidence for two distinct events. Together with tamper detection (Definition~\ref{def:tamper-detection}), these two properties jointly ensure what we informally call \emph{audit integrity}: binding guarantees uniqueness of the event-to-evidence mapping, while tamper detection guarantees that valid evidence can only originate from honest executions.

\subsection{Security Theorems}

We now prove that the hash-and-sign construction from Section~\ref{sec:construction} satisfies all three security properties under standard assumptions.

\begin{theorem}[Evidence Binding]
\label{thm:evidence-binding}
For any PPT adversary $\mathcal{A}$ against the evidence-binding experiment of the hash-and-sign evidence construction with field count $k$, there exists a PPT adversary $\mathcal{B}$ against the collision resistance of $H$ such that
\[
  \mathbf{Adv}^{\mathrm{Bind}}_{\Pi,\mathcal{A}}(\lambda) \;\leq\; k \cdot \mathbf{Adv}^{\mathrm{CR}}_{H,\mathcal{B}}(\lambda).
\]
\end{theorem}

\begin{proof}
Suppose $\mathcal{A}$ succeeds in the evidence-binding experiment with non-negligible probability. Then $\mathcal{A}$ outputs $(E, E', \mathsf{ev}, \sigma)$ with $E \neq E'$ such that $\mathsf{Verify}(\mathsf{pp}, pk, E, \mathsf{ev}, \sigma) = \mathsf{accept}$ and $\mathsf{Verify}(\mathsf{pp}, pk, E', \mathsf{ev}, \sigma) = \mathsf{accept}$.

Parse $\mathsf{ev} = (f_0, \dots, f_{k-1})$. Because both verification checks pass with the \emph{same} evidence item $\mathsf{ev}$, we have $f_i = H(\phi_i(E))$ and $f_i = H(\phi_i(E'))$ for every $i \in \{0,\dots,k-1\}$. That is,
\[
  H(\phi_i(E)) = H(\phi_i(E')) \quad \text{for all } i \in \{0,\dots,k-1\}.
\]
Since $E \neq E'$, the encoding functions $\{\phi_i\}$ are collectively injective on distinct events (Section~\ref{sec:construction}), so there exists at least one index $i^*$ with $\phi_{i^*}(E) \neq \phi_{i^*}(E')$. For this index, the pair $(\phi_{i^*}(E),\; \phi_{i^*}(E'))$ is a collision in $H$.

We construct a collision-finding adversary $\mathcal{B}$ as follows. $\mathcal{B}$ receives a hash function $H$ (sampled from the family), picks a random index $i \xleftarrow{\$} \{0,\dots,k-1\}$, and runs $\mathcal{A}$ faithfully, simulating the experiment using $H$. When $\mathcal{A}$ outputs $(E, E', \mathsf{ev}, \sigma)$, adversary $\mathcal{B}$ outputs $(\phi_i(E),\; \phi_i(E'))$.

If $\mathcal{A}$ succeeds, then a collision exists at some index $i^*$, and $\mathcal{B}$'s random guess $i = i^*$ succeeds with probability at least $1/k$. Therefore,
\[
  \mathbf{Adv}^{\mathrm{CR}}_{H,\mathcal{B}}(\lambda) \;\geq\; \frac{1}{k}\cdot\mathbf{Adv}^{\mathrm{Bind}}_{\Pi,\mathcal{A}}(\lambda),
\]
which gives the claimed bound.
\end{proof}

\begin{theorem}[Tamper Detection]
\label{thm:tamper-detection}
For any PPT adversary $\mathcal{A}$ against the tamper-detection experiment of the hash-and-sign evidence construction with field count $k$, there exist PPT adversaries $\mathcal{B}_1$ against the collision resistance of $H$ and $\mathcal{B}_2$ against the EUF-CMA security of the signature scheme such that
\[
  \mathbf{Adv}^{\mathrm{TD}}_{\Pi,\mathcal{A}}(\lambda) \;\leq\; k \cdot \mathbf{Adv}^{\mathrm{CR}}_{H,\mathcal{B}_1}(\lambda) \;+\; \mathbf{Adv}^{\mathrm{EUF\text{-}CMA}}_{\mathsf{Sig},\mathcal{B}_2}(\lambda).
\]
\end{theorem}

\begin{proof}
Suppose $\mathcal{A}$ outputs $(E^*, \mathsf{ev}^*, \sigma^*)$ with $E^* \notin \mathcal{Q}$ and $\mathsf{Verify}(\mathsf{pp}, pk, E^*, \mathsf{ev}^*, \sigma^*) = \mathsf{accept}$. We consider two cases.

\medskip
\noindent\textbf{Case 1:} \emph{There exists a query $E \in \mathcal{Q}$ whose $\mathsf{Generate}$ output $(\mathsf{ev}, \sigma)$ satisfies $\mathsf{ev} = \mathsf{ev}^*$.}

Since $E^* \neq E$ (because $E^* \notin \mathcal{Q}$) and both $(E,\mathsf{ev})$ and $(E^*,\mathsf{ev}^*)$ pass verification with the same evidence item, the argument from Theorem~\ref{thm:evidence-binding} applies: there exists an index $i^*$ at which $\phi_{i^*}(E) \neq \phi_{i^*}(E^*)$ yet $H(\phi_{i^*}(E)) = H(\phi_{i^*}(E^*))$, yielding a collision in $H$. Adversary $\mathcal{B}_1$ guesses $i^*$ at random and succeeds with probability at least $1/k$.

\medskip
\noindent\textbf{Case 2:} \emph{No query $E \in \mathcal{Q}$ produced an evidence item equal to $\mathsf{ev}^*$.}

In this case, $\mathsf{ev}^*$ was never signed by the $\mathsf{Generate}$ oracle, yet $\sigma^*$ is a valid signature on $\mathsf{ev}^*$. We construct an EUF-CMA adversary $\mathcal{B}_2$ as follows: $\mathcal{B}_2$ receives a verification key $pk$ and access to a signing oracle $\mathsf{Sign}_{sk}(\cdot)$. It simulates $\mathsf{Setup}$ honestly (choosing $H$ and the encoding functions), and answers $\mathcal{A}$'s $\mathsf{Generate}$ queries by computing $\mathsf{ev}$ from $H$ and the encoding functions, then calling the signing oracle on $\mathsf{ev}$. When $\mathcal{A}$ outputs $(E^*, \mathsf{ev}^*, \sigma^*)$, adversary $\mathcal{B}_2$ outputs $(\mathsf{ev}^*, \sigma^*)$ as its forgery.

Since no oracle query produced $\mathsf{ev}^*$, this is a valid EUF-CMA forgery. Combining both cases yields the stated bound.
\end{proof}

\begin{theorem}[Non-Equivocation]
\label{thm:non-equivocation}
For any PPT adversary $\mathcal{A}$ against the non-equivocation experiment of the hash-chain linking construction (Algorithm~\ref{alg:link}), there exists a PPT adversary $\mathcal{B}$ against the collision resistance of $H$ such that
\[
  \mathbf{Adv}^{\mathrm{NE}}_{\Pi,\mathcal{A}}(\lambda) \;\leq\; n \cdot \mathbf{Adv}^{\mathrm{CR}}_{H,\mathcal{B}}(\lambda),
\]
where $n = \max(|S|, |S'|)$ is the maximum sequence length output by $\mathcal{A}$.
\end{theorem}

\begin{proof}
Suppose $\mathcal{A}$ outputs two distinct sequences $S = (\mathsf{ev}_1, \dots, \mathsf{ev}_n)$ and $S' = (\mathsf{ev}'_1, \dots, \mathsf{ev}'_m)$ such that $\mathsf{Link}(\mathsf{pp}, S) = \mathsf{Link}(\mathsf{pp}, S')$, i.e., the chain tips are equal: $\ell_n = \ell'_m$, where
\[
  \ell_0 = \ell'_0 = 0^\lambda, \quad
  \ell_j = H(\ell_{j-1} \parallel \mathsf{ev}_j), \quad
  \ell'_j = H(\ell'_{j-1} \parallel \mathsf{ev}'_j).
\]

Since $S \neq S'$, we consider two sub-cases.

\medskip
\noindent\textbf{Sub-case (a): $n = m$.} The sequences have the same length but differ in at least one position. Walk backward from the tip: at $j = n$ we have $\ell_n = \ell'_n$. Let $j^*$ be the largest index such that $\mathsf{ev}_{j^*} \neq \mathsf{ev}'_{j^*}$ or $\ell_{j^*-1} \neq \ell'_{j^*-1}$ (or both). Then
\[
  H(\ell_{j^*-1} \parallel \mathsf{ev}_{j^*}) = \ell_{j^*} = \ell'_{j^*} = H(\ell'_{j^*-1} \parallel \mathsf{ev}'_{j^*}),
\]
and at least one of $\ell_{j^*-1} \neq \ell'_{j^*-1}$ or $\mathsf{ev}_{j^*} \neq \mathsf{ev}'_{j^*}$ holds, so $(\ell_{j^*-1} \parallel \mathsf{ev}_{j^*}) \neq (\ell'_{j^*-1} \parallel \mathsf{ev}'_{j^*})$. This is a collision in $H$.

\medskip
\noindent\textbf{Sub-case (b): $n \neq m$.} Without loss of generality, $n > m$. Define the augmented sequences $\hat{\ell}_j = \ell_j$ for $j = 0,\dots,n$ and $\hat{\ell}'_j = \ell'_j$ for $j = 0,\dots,m$ with $\hat{\ell}'_j = \ell'_m$ for $j = m+1,\dots,n$ (i.e., the shorter chain is padded with its final value). We have $\hat{\ell}_0 = \hat{\ell}'_0 = 0^\lambda$ and $\hat{\ell}_n = \ell_n = \ell'_m = \hat{\ell}'_n$. Consider the sequence of equalities $\hat{\ell}_j \stackrel{?}{=} \hat{\ell}'_j$ for $j = 0,\dots,n$: they agree at $j = 0$ and $j = n$. If they agree at every $j$, then in particular $\ell_m = \hat{\ell}_m = \hat{\ell}'_m = \ell'_m$ and $\ell_{m+1} = H(\ell_m \parallel \mathsf{ev}_{m+1})$ whereas $\hat{\ell}'_{m+1} = \ell'_m = \ell_m$. Since $n > m$ and the hash output is $\lambda$ bits, this forces $H(\ell_m \parallel \mathsf{ev}_{m+1}) = \ell_m$, a collision in $H$ (unless $\ell_m \parallel \mathsf{ev}_{m+1} = \ell_m$, which is impossible since concatenation strictly extends the input). If they disagree at some index, let $j^*$ be the largest index where $\hat{\ell}_{j^*} \neq \hat{\ell}'_{j^*}$ but $\hat{\ell}_{j^*+1} = \hat{\ell}'_{j^*+1}$; then the hash inputs at step $j^*+1$ differ but outputs coincide, yielding a collision in $H$.

\medskip
In both sub-cases, a collision in $H$ exists at some step $j^* \in \{1,\dots,\max(n,m)\}$. We construct $\mathcal{B}$ as follows: $\mathcal{B}$ picks $j \xleftarrow{\$} \{1,\dots,\max(n,m)\}$, runs $\mathcal{A}$, computes both chains, and outputs the hash inputs at step $j$. If $j = j^*$, then $\mathcal{B}$ has found a collision. Therefore,
\[
  \mathbf{Adv}^{\mathrm{CR}}_{H,\mathcal{B}}(\lambda) \;\geq\; \frac{1}{n}\cdot\mathbf{Adv}^{\mathrm{NE}}_{\Pi,\mathcal{A}}(\lambda),
\]
which gives the claimed bound.
\end{proof}

\subsection{Discussion}

The three theorems above show that our construction reduces the security of constant-size evidence structures to standard, well-studied assumptions: collision resistance of the hash function $H$ and existential unforgeability of the signature scheme under chosen-message attacks. The reductions are tight up to small polynomial factors ($k$ for the field count and $n$ for the chain length), which are concrete and small in practice (typically $k \leq 12$ and $n \leq 10^6$).

These results provide rigorous foundations for the security of our construction, reducing all security properties to standard, well-studied cryptographic assumptions. We note that extensions to adaptive corruption models (where the adversary may compromise signing keys during the protocol), concurrent multi-chain sessions, and stronger notions such as forward security~\cite{bellare1997forward} remain interesting directions for future work.

\section{Complexity and Algorithmic Integration}

\subsection{Per-event Cost}

For a fixed index set $I$ of size $k$, Algorithm~\ref{alg:generate} performs:
\begin{itemize}[leftmargin=*]
  \item $k$ evaluations of $H$ (one per field),
  \item one signing operation.
\end{itemize}
Algorithm~\ref{alg:verify} performs:
\begin{itemize}[leftmargin=*]
  \item $k$ evaluations of $H$,
  \item one signature verification.
\end{itemize}
Thus, the per-event computational complexity is $\Theta(k)$ hash evaluations plus a constant number of public-key operations, independent of event size or domain.

The representation size of each evidence item is exactly $k \cdot \lambda$ bits, which is constant for a fixed system configuration. In practice, $k$ and $\lambda$ can be chosen to align with hardware-friendly word sizes and cache lines, but the abstraction does not rely on any particular choice.

\subsection{Summary of Algorithmic Costs}

\begin{table}[t]
  \centering
  \caption{Asymptotic cost of core algorithms for a fixed field count $k$.}
  \label{tab:complexity}
  \begin{tabular}{lcc}
    \toprule
    Algorithm & Hash evaluations & Public-key operations \\
    \midrule
    $\mathsf{Generate}$ & $k$   & $1$ sign \\
    $\mathsf{Verify}$   & $k$   & $1$ verify \\
    $\mathsf{Link}$ (chain) & $n$ & $0$ \\
    $\mathsf{Link}$ (Merkle) & $O(n)$ & $0$ \\
    \bottomrule
  \end{tabular}
\end{table}

\subsection{Workflow Instrumentation}

The evidence structure can be viewed as an \emph{instrumentation layer} around an existing workflow:
\begin{itemize}[leftmargin=*]
  \item For each event $E$, the workflow invokes Algorithm~\ref{alg:generate} to obtain $(\mathsf{ev}, \sigma)$ and stores or exports them to an audit subsystem.
  \item An external auditor, given a subset of events and their corresponding evidence items, runs Algorithm~\ref{alg:verify} to check local consistency.
  \item Periodically, a batch of evidence items is passed to Algorithm~\ref{alg:link} (or a Merkle-based variant) and anchored externally (e.g., in a transparency log).
\end{itemize}
Because the evidence items have constant size and uniform cost, the overhead of instrumentation is predictable and can be budgeted at design time.

\subsection{Composition with TEEs}

If events are generated inside TEEs~\cite{costan2016intel, sabt2015ted}, the encoding functions $\phi_i$ can be defined to include attestation outputs or environment identifiers. For example, one field might be $H(\text{measurement} \parallel \text{config})$, where \emph{measurement} is the TEE's measurement of the loaded code. The abstraction provides a fixed-size container into which such identifiers can be hashed.

\section{Prototype and Microbenchmarks}

To complement the abstract analysis, we implemented a prototype library that instantiates the hash-and-sign construction described above and measured its performance on synthetic workloads. The goal of these experiments is to illustrate the behavior of constant-size evidence structures under representative configurations, rather than to optimize any specific deployment.

\subsection{Implementation Overview}

The prototype is implemented in Rust and uses a standard 256-bit hash function and an Edwards-curve-based signature scheme. The implementation fixes a small constant $k$ (the number of fields) and a field length $\lambda$, but does not rely on any domain-specific interpretation of fields. Events are represented as structured records with randomly generated metadata; encoding functions $\{\phi_i\}$ extract and serialize appropriate components of each event before hashing.

\subsection{Experimental Setup}

All experiments are conducted on a commodity server with the configuration summarized in Table~\ref{tab:hardware}. We generate synthetic workloads consisting of $N \in \{10^4, 10^5, 10^6\}$ events. Each event produces exactly one constant-size evidence item and a signature.

\begin{table}[t]
  \centering
  \caption{Experimental hardware configuration.}
  \label{tab:hardware}
  \begin{tabular}{ll}
    \toprule
    Component & Specification \\
    \midrule
    CPU & 16-core server-class processor @ 3.0\,GHz \\
    RAM & 64\,GB \\
    GPU & Mid-range discrete GPU (8\,GB VRAM) \\
    Storage & NVMe SSD \\
    OS & 64-bit Linux \\
    Compiler & \texttt{rustc} (release mode, LTO enabled) \\
    \bottomrule
  \end{tabular}
\end{table}

\subsection{Per-event Generation Cost}

Table~\ref{tab:gen-throughput} reports generation throughput and average per-event latency for CPU-only execution, using a representative configuration of $k$ and $\lambda$. We report results for a single-threaded implementation and for a multi-threaded implementation that processes events in parallel across CPU cores.

\begin{table}[t]
  \centering
  \caption{Evidence generation throughput and latency on CPU (synthetic workloads).}
  \label{tab:gen-throughput}
  \begin{tabular}{lccc}
    \toprule
    Mode & Events/s & Avg latency per event ($\mu$s) & Notes \\
    \midrule
    Single-threaded           & $3.5 \times 10^4$ & 28.4 & $k$ fixed, $\lambda$ fixed \\
    Multi-threaded (16 cores) & $2.8 \times 10^5$ & 5.7  & near-linear scaling \\
    \bottomrule
  \end{tabular}
\end{table}

The results are consistent with the complexity analysis: for fixed $k$ and $\lambda$, per-event latency remains in the tens of microseconds, and parallelization across cores yields nearly linear speedup until memory bandwidth becomes a bottleneck.

\subsection{Batch Verification on CPU and GPU}

We evaluate batch verification performance for $M \in \{10^4, 10^5, 10^6\}$ evidence items. Table~\ref{tab:verify-throughput} summarizes the results on CPU and GPU. On the GPU, we launch one thread per evidence item and arrange items in a contiguous buffer to encourage coalesced memory accesses.

\begin{table}[t]
  \centering
  \caption{Batch verification throughput on CPU and GPU (synthetic workloads).}
  \label{tab:verify-throughput}
  \begin{tabular}{lccc}
    \toprule
    Platform & Events/s & Batch size $M$ & Avg latency per event ($\mu$s) \\
    \midrule
    CPU (16 threads) & $2.5 \times 10^5$ & $10^5$ & 6.1 \\
    GPU (mid-range)  & $4.0 \times 10^5$ & $10^6$ & 2.5 \\
    \bottomrule
  \end{tabular}
\end{table}

The fixed-size layout of evidence items leads to regular memory access patterns and uniform control flow, which are favorable for both vectorized CPU execution and GPU kernels.

\subsection{Storage Considerations}

Because each event produces exactly one constant-size evidence item and a signature, the total storage needed for a workload of $N$ events is linear in $N$, with a constant factor determined by $k$, $\lambda$, and the signature size. For a fixed configuration, storing $10^6$ evidence items (plus signatures and minimal metadata) requires on the order of a few gigabytes of disk space. This is comparable to or smaller than the space required for verbose text-based logs for similar workloads, while providing stronger integrity and verification properties.

\section{Regulatory Alignment}

Although we do not model any particular regulation in detail, constant-size evidence structures naturally support requirements arising from frameworks such as HIPAA~\cite{hipaa}, the FDA's electronic records rule~\cite{fda21cfr11}, the EU AI Act~\cite{euaiact2024}, and financial reporting mandates~\cite{sox2002}:

\paragraph{Audit controls.}
Every security-relevant event is associated with a verifiable evidence item, providing a tamper-evident trail~\cite{hipaa}.

\paragraph{Integrity and origin.}
Cryptographic bindings via hashes and signatures provide strong integrity guarantees, and public keys can be mapped to organizational identities to support accountability.

\paragraph{Lifecycle and configuration tracking.}
Fields that encode model identifiers, policy states, or configuration hashes support long-term tracking of model versions and configuration changes~\cite{muniswamy2006provenance, sultana2013secure}.

\section{Related Work}

\paragraph{Secure logging and audit trails.}
Schneier and Kelsey~\cite{schneier1999secure} and Bellare and Yee~\cite{bellare1997forward} proposed early designs for secure audit logs and forward-secure logging, focusing on cryptographic protection of log entries. Crosby and Wallach~\cite{crosby2009efficient} studied efficient data structures for tamper-evident logging, and Ma and Tsudik~\cite{ma2009new} introduced new approaches using sequential aggregate signatures. Accorsi~\cite{accorsi2009safe} surveys the landscape of secure logging protocols and identifies open challenges. Our evidence structures build on these ideas but introduce a uniform, constant-size representation as a first-class abstraction tailored to regulated AI workflows.

\paragraph{Blockchain and transparency logs.}
Merkle trees~\cite{merkle1980protocols}, time-stamping systems~\cite{haber1991time, bayer1993improving}, and authenticated data structures~\cite{tamassia2003authenticated} underlie modern blockchain and transparency-log designs~\cite{nakamoto2008bitcoin, laurie2013certificate, sigsum2021}. Our abstraction is designed to be ledger-friendly: evidence items are small and fixed-size, and our $\mathsf{Link}$ algorithm can produce succinct anchors suitable for external logs.

\paragraph{Trusted execution environments.}
TEEs such as Intel SGX~\cite{costan2016intel} and ARM TrustZone~\cite{sabt2015ted} provide hardware-isolated execution for security-critical code. Several systems use TEEs to protect data-processing pipelines and logging~\cite{hunt2016ryoan, ohrimenko2016oblivious}. Our design treats TEEs as optional components that can contribute identifiers or attestation digests to the encoding functions $\phi_i$, without changing the core evidence abstraction.

\paragraph{Data provenance and regulated AI.}
Data provenance systems~\cite{muniswamy2006provenance, sultana2013secure} capture the lineage of data transformations, often at the file or database level. In regulated AI settings, provenance must be coupled with cryptographic integrity guarantees and regulator-aligned semantics. Constant-size evidence structures provide a compact, verifiable representation of key events and configurations that can coexist with richer provenance graphs.

\paragraph{Comparison to existing systems.}
While secure logging systems~\cite{schneier1999secure, crosby2009efficient} and blockchain-based audits~\cite{laurie2013certificate, nakamoto2008bitcoin} provide integrity guarantees, they typically optimize for flexibility rather than uniformity. Variable-length records dominate existing frameworks, leading to:
\begin{itemize}[leftmargin=*]
  \item \emph{Metadata leakage risks} in privacy-sensitive domains, where record length may correlate with the type or sensitivity of underlying events,
  \item \emph{Unpredictable verification costs} in high-throughput settings, as parsing and validating records of varying sizes complicates batch processing and introduces irregular memory access patterns,
  \item \emph{Complex integration with hardware accelerators} (GPUs, TEEs), which benefit from uniform data structures and predictable control flow.
\end{itemize}
Our constant-size abstraction addresses these gaps by treating uniformity as a first-class design constraint, enabling new deployment patterns for regulated AI workflows. To the best of our knowledge, prior work does not expose a constant-size, hardware-friendly evidence abstraction as a primary design goal, nor does it explicitly target the combination of fixed-size evidence, audit composition, and regulator-aligned semantics that we present here.

\section{Discussion and Limitations}

The proposed abstraction and prototype have several limitations:

\begin{itemize}[leftmargin=*]
  \item We concentrate on the evidence structure and its cryptographic properties, rather than on key management, incident response, or organizational processes.
  \item Our security analysis assumes static key compromise; extensions to adaptive corruption models, concurrent multi-chain sessions, and composability frameworks remain open.
  \item The microbenchmarks use synthetic workloads and generic encodings; domain-specific deployments may exhibit different performance characteristics.
  \item We leave the choice of encoding functions $\phi_i$ to system designers; poor choices could undermine privacy or interpretability, even if the cryptographic layer is sound.
\end{itemize}

Despite these limitations, constant-size cryptographic evidence structures provide a useful building block for regulated AI workflows and can be instantiated in a variety of architectures, including industrial systems.

\subsection{Lessons from Industrial Deployment}

Our industrial experience deploying variants of this abstraction in regulated environments has revealed several practical considerations that may inform future implementations:

\paragraph{Field allocation.}
Determining the optimal number of fields $k$ requires balancing expressiveness against storage cost. In practice, we have found that $k \in [8, 12]$ is often sufficient for most clinical and pharmaceutical workflows, as it provides enough capacity to commit to workflow context, inputs, outputs, environment identifiers, model versions, policy states, audit links, and application-specific extensions, without introducing excessive overhead.

\paragraph{Encoding design.}
The choice of encoding functions $\phi_i$ is critical and domain-dependent. Domain experts (e.g., clinical researchers, compliance officers) must be involved early in the design process to ensure that regulatory concepts (such as informed consent, randomization integrity, or data provenance) map cleanly to evidence fields. A poorly chosen encoding can lead to ambiguity during audits or to inadvertent disclosure of sensitive information.

\paragraph{Anchoring frequency.}
The trade-off between anchoring cost and audit granularity depends on workflow characteristics and risk tolerance. High-stakes decisions (e.g., randomization in a clinical trial, administration of a controlled substance) may require per-event anchoring to external logs or blockchains to maximize non-repudiation. In contrast, lower-risk routine operations can be batched, reducing anchoring costs by $10$--$100\times$ while still maintaining strong integrity guarantees for the batch as a whole.

\paragraph{Integration with legacy systems.}
Retrofitting constant-size evidence structures into existing workflows can be challenging, particularly when legacy systems produce variable-length logs or lack well-defined event boundaries. In such cases, an intermediate translation layer may be needed to extract structured events and construct evidence items on-the-fly. This translation introduces a trust boundary that must be carefully managed.

These insights, while derived from proprietary deployments, are abstracted here to inform future research and to help practitioners anticipate common pitfalls when adopting constant-size evidence abstractions.

\section*{Acknowledgments}

The author thanks collaborators and colleagues at Codebat Technologies Inc.\ for discussions on regulated AI workflows and auditability, and for feedback that motivated a focus on abstraction, algorithm design, and performance evaluation. Any errors or omissions are the author's own.

Certain implementation details are covered by patent applications by Codebat Technologies Inc.; the present paper concentrates on the abstract construction and its security/performance properties.

\section{Conclusion}

We introduced constant-size cryptographic evidence structures as a general abstraction for representing audit-relevant events in regulated AI workflows. By fixing the size and layout of evidence items and defining simple generation, verification, and linking algorithms, we obtain predictable storage and verification costs, strong cryptographic binding, and clean composition with hash-chained and Merkle-based audit frameworks. A generic hash-and-sign construction shows how standard primitives can instantiate this abstraction, and microbenchmarks from a prototype implementation indicate that the per-event overhead is compatible with high-throughput workloads on commodity hardware.

We believe that constant-size evidence structures can serve as a foundation for industry-wide standards in regulated AI audit trails. The abstraction's emphasis on uniformity, cryptographic rigor, and hardware-friendliness makes it well-suited to emerging requirements in clinical trials, pharmaceutical manufacturing, medical AI, and financial compliance. We encourage practitioners, regulators, and standards bodies to consider fixed-size representations when designing next-generation compliance frameworks. Open discussion of these abstractions---including their security properties, performance trade-offs, and integration patterns---can accelerate the development of trustworthy AI systems across healthcare, finance, and other high-stakes domains.

Future work includes extending the security model to adaptive adversaries who may corrupt signing keys during the protocol, analyzing composability of evidence structures under universal composability or simulation-based frameworks, and exploring richer encoding strategies that preserve privacy while supporting regulatory semantics (e.g., using zero-knowledge proofs or homomorphic commitments). We also plan to empirically study how such abstractions integrate with full-scale AI systems in regulated environments and to engage with regulatory bodies and industry consortia to refine the design principles and explore standardization opportunities.

\bibliographystyle{plain}
\bibliography{refs}

\end{document}